\documentclass[%
 preprint,
 amsmath,amssymb,
 aps,
]{revtex4-2}

\usepackage{graphicx}
\usepackage{dcolumn}
\usepackage{bm}
\usepackage{hyperref}
\usepackage{color}

\begin{document}


\title{Rapid chemically selective 3D imaging in the mid-infrared with a Si-based camera}

\author{Eric O. Potma$^{1,2}$\email{epotma@uci.edu}, David Knez$^1$, Yong Chen$^3$, Amanda Durkin$^{2}$, Alexander Fast$^{2}$, Mihaela Balu$^{2}$, Brenna Norton-Baker$^1$, Tommaso Baldacchini$^1$, Rachel Martin$^1$, Dmitry A. Fishman$^1$\email{dmitryf@uci.edu}}
\affiliation{$^1$Department of Chemistry, University of California, Irvine, CA 92697, USA}
\affiliation{$^2$Beckman Laser Institute, University of California, Irvine, CA 92697, USA}
\affiliation{$^3$Epstein Department of Industrial and Systems Engineering, Los Angeles, CA 90089, USA}

\date{\today}

\begin{abstract}
The emerging technique of mid-infrared optical coherence tomography (MIR-OCT) takes advantage of the reduced scattering of MIR light in various materials and devices, enabling tomographic imaging at deeper penetration depths. Because of challenges in MIR detection technology, the image acquisition time is however significantly longer than for tomographic imaging methods in the visible/near-infrared. Here we demonstrate an alternative approach to MIR tomography with high-speed imaging capabilities. Through femtosecond non-degenerate two-photon absorption of MIR light in a conventional Si-based CCD camera, we achieve wide-field, high-definition tomographic imaging with chemical selectivity of structured materials and biological samples in mere seconds.
\end{abstract}

\maketitle


\section{\label{sec:first}Introduction}
Supplementing optical imaging with spectroscopic information enables the identification of objects based not only on their morphology but also on their chemical composition. The mid-infrared (MIR) region of the electromagnetic spectrum, which spans the 2~$\mu$m – 10~$\mu$m wavelength range, has been of particular interest. This range matches the energies of the fundamental vibrations of chemical bonds and moieties, and, therefore, constitutes a premier window for spectroscopic imaging.\cite{Bhargava2012,Wetzel1999} For this reason, MIR radiation has been used for target discrimination in stand-off detection mode.\cite{Wang2005} In addition, when combined with interferometric gating, MIR light has proven attractive for tomographic imaging, allowing the visualization and three-dimensional reconstruction of a variety of structured materials. In particular, compared to visible and near-IR (NIR) radiation, MIR light has a much higher penetration depth in highly scattering materials such as ceramics, paints, and printed electronics, which has prompted the development of MIR-based optical coherence tomography (MIR-OCT) techniques.\cite{Cheung2015,Colley2007,Guilhaumou1998,Ishida2012,Jamme2013,Martin2013,Sharma2008,Zorin2020,Zorin2018}. 

Although the unique imaging capabilities of MIR tomographic imaging, and MIR-OCT in particular, address an important need in the characterization of structured materials, its practical implementation is hampered by technical hurdles. For instance, fast and low-noise detection of MIR radiation, a prerequisite for rapid imaging, remains a challenge for existing detector technologies. MIR detectors, such as those based on low bandgap materials like InSb and HgCdTe, suffer from a high thermal background. Moreover, MIR detector arrays typically have fewer elements compared to their visible/NIR counterparts, thus limiting high-definition imaging capabilities. These obstacles have spurred many developments that aim to convert the information encoded in MIR light into vis/NIR radiation.\cite{Bai2019,Cirloganu2010,Cirloganu2011,Junaid2018,Kviatkovsky2020,Paterova2020} Such spectral conversion enables the use of mature detector technology based on Si or other wide bandgap semiconductor materials.\cite{Barh2019,Fishman2011,Johnson2012,Junaid2019,Knez2020,Hanninen2019,Hanninen2018} This strategy has been leveraged to improve the performance of MIR OCT, using either nonlinear up-conversion\cite{Israelsen2019} or nonlinear interferometry with entangled photons.\cite{Vanselow2020,Rojas2020,Machado2020,Paterova2018,Valles2018} MIR tomographic images have been recorded at sub-10~$\mu$m axial resolution and total acquisition times of minutes per volume.

Despite these important advances in MIR-OCT, the total acquisition time for volumetric images is still rather long and relies on lateral raster scanning of the beam, rendering current approaches less practical for time-sensitive applications. Furthermore, signal levels appear insufficient for examining weakly reflective interfaces of organic materials nor do current OCT applications take clear advantage of the spectroscopic sensitivity afforded by MIR light. Here we develop a new, high-speed 3D imaging technique that overcomes these shortcomings in MIR tomographic imaging. Instead of relying on interferometric gating to achieve depth resolution, our approach uses a nonlinear optical gate provided by an additional femtosecond pulse through the process of non-degenerate two-photon absorption (NTA) in a wide bandgap semiconducting photodetector. This effect has been discussed in depth\cite{Cirloganu2011,Fishman2011,Bristow2007,Cox2019,Hutchings1992,Fang2020} and the possibility of using NTA for rapid 2D mid-IR imaging has been demonstrated.\cite{Knez2020} In fact, the principle of NTA has recently been used to acquire tomographic images in the MIR range with the aid of a single pixel GaN photodiode.\cite{Pattanaik2016} To achieve 3D imaging, the object was raster scanned across a focused MIR beam, requiring multiple minutes to build up a volumetric dataset when using lock-in detection. In contrast, we apply this principle in a massively parallel fashion through the use of a 1.4 Mpx Si CCD camera, omitting the need for lateral scanning altogether and enabling the acquisition of 3D images in mere seconds or faster - acquisition rates that are up to two orders of magnitude higher than the present standard. This novel detection strategy permits 3D MIR imaging at high sensitivity. We show that NTA-enabled tomography allows background-free 3D MIR imaging of weakly reflective interfaces of organic materials, objects underneath a 3~mm thick GaAs wafer and even targets hidden under a 190 $\mu$m layer of water, using MIR illumination doses as low as 0.4~mW/cm$^2$. To emphasize the chemical selectivity of MIR tomography, we demonstrate 3D images of polymer structures and protein crystals with spectroscopic contrast based on fundamental vibrational transitions in the 2000-3000 cm$^{-1}$ range.
\begin{figure}[ht]
    \centering
    \includegraphics[width=14cm]{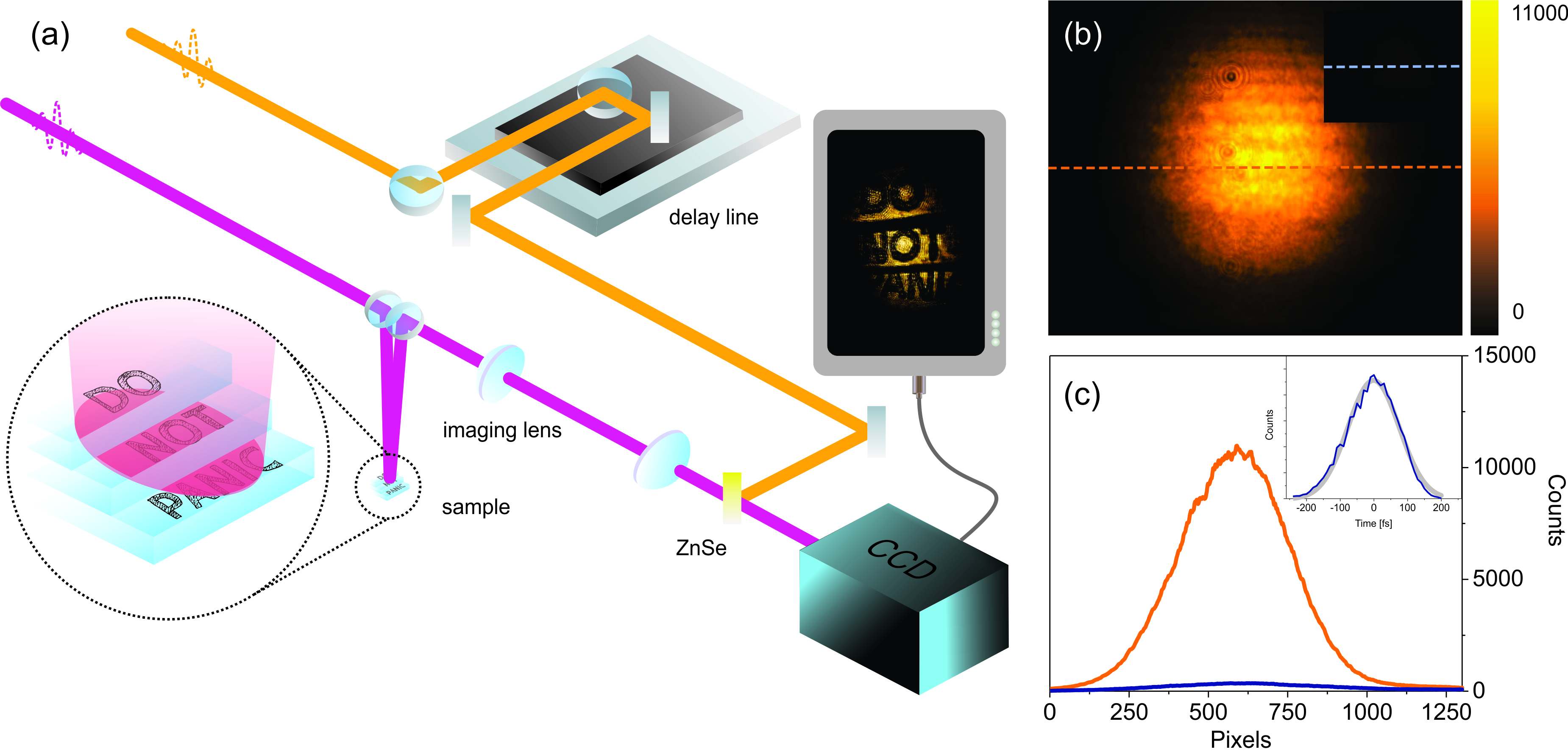}
    \caption{(a) Schematic representation of setup. (b) Beam image on CCD Si chip with MIR (2850~cm$^{-1}$, 3500~nm) and gate NIR pulses (8333~cm$^{-1}$, 1200~nm). Inset shows gate pulse only image at the same scale. (c) Spatial cross-section of beam image on CCD Si chip with and without MIR pulse. Inset: temporal cross-correlation of MIR and gate pulse, indicating a 110~fs pulse width (gray line – Gaussian fit).}
    \label{fig:fig1}
\end{figure}

\section{\label{sec:second}Rapid MIR 3D imaging}
Our experimental concept is schematically depicted in Figure \ref{fig:fig1}a. Two 100~fs pulse trains, a tunable MIR beam and a fixed 1200~nm (8333~cm$^{-1}$) NIR gate are spatially overlapped on the Si chip of a CCD camera. The MIR beam path passes onto the sample and the reflected/scattered light is collected by a 100~mm CaF$_2$ lens that projects an image of the sample onto the CCD in a 1:1 fashion (effective NA~0.015). An NTA signal is generated in the Si chip whenever the MIR temporally overlaps with the NIR gate pulse. The temporal gate, determined by the cross correlation of the pulses, selectively registers MIR light that has traveled a pre-set path length. This allows depth-dependent detection of reflected/scattered light off the illuminated sample interface. In this wide-field imaging geometry, acquisition of a full 3D scan requires a single axis scan of the time delay. In principle, this strategy enables 3D imaging with volumetric acquisition rates limited only by the frame rate of the CCD camera.

Figure \ref{fig:fig1}b shows an image of the MIR beam profile (2850~cm$^{-1}$, 3500~nm) acquired through NTA. Compared to previous reports based on picosecond pulses\cite{Knez2020}, the increased irradiance of femtosecond pulses used here enables more efficient NTA detection. The intensity ratio between the MIR and the gate pulse can be flexibly chosen to raise the NTA signal relative to the degenerate two-photon absorption (DTA) background induced by the gate pulse.\cite{Cirloganu2011,Fishman2011,Knez2020,Fang2020} This is highlighted in the inset of Figure \ref{fig:fig1}b, which shows the DTA background signal when the Si chip is exposed only to the gate pulse. The NTA to DTA background signal is 34dB (standard deviation $\sim2$\%, 100~ms per frame, \href{https://www.chem.uci.edu/~dmitryf/images/Supplementary Information.pdf}{{\color{blue}Supplementary}} Figure SF1) at the center of the beam profile, with a ratio up to 40~dB towards the wings of the intensity distribution. The high signal to background ratio permits imaging without modulation/demodulation techniques, beam profile/intensity pre-characterization or post-processing of the images.

Other than in MIR OCT, where the axial resolution is determined by the spectral width of the light source via interferometric gating, in NTA tomographic imaging the resolution is determined by the temporal cross correlation of the MIR and gate pulses. The inset of Figure \ref{fig:fig1}c shows the cross-correlation of pulses at the camera chip, indicating a MIR pulse width of $\sim110$~fs (FWHM) assuming a Gaussian pulse shape. This temporal pulse width corresponds to an axial resolution of $\sim15~\mu$m (FWHM) or 12.7~$\mu$m (1/2e$^2$) in vacuum (\href{https://www.chem.uci.edu/~dmitryf/images/Supplementary Information.pdf}{{\color{blue}Supplementary}} Figure SF5). Note that the axial resolution is much higher than the resolution set by the Rayleigh range of the imaging system for the MIR beam (NA=0.015, $z_R > 3$~mm), and is comparable to the confocal resolution offered by a $\rm{NA}>0.65$ lens.

We first perform 2D imaging to illustrate the image quality of the technique. In Supplementary  \href{https://www.chem.uci.edu/~dmitryf/images/VideoV5.avi}{{\color{blue}Video 1}}, we show a live recording of Tetramorium caespitum (pavement ant) moving freely on a glass microscope slide. At the low MIR illumination levels of 1~mW/cm$^2$, the ant is active and appears unaffected by the radiation. It is interesting to observe that the chitin exoskeleton of the legs appears darker when the MIR is tuned to 2850 cm$^{-1}$, while it appears semi-transparent when tuned to 2450 cm$^{-1}$. The different contrast reflects the differences in MIR absorption of chitin, a polysaccharide rich in CH2 groups (\href{https://www.chem.uci.edu/~dmitryf/images/Supplementary Information.pdf}{{\color{blue}Supplementary}} Figure SF2). 
\begin{figure}
    \centering
    \includegraphics[width=14cm]{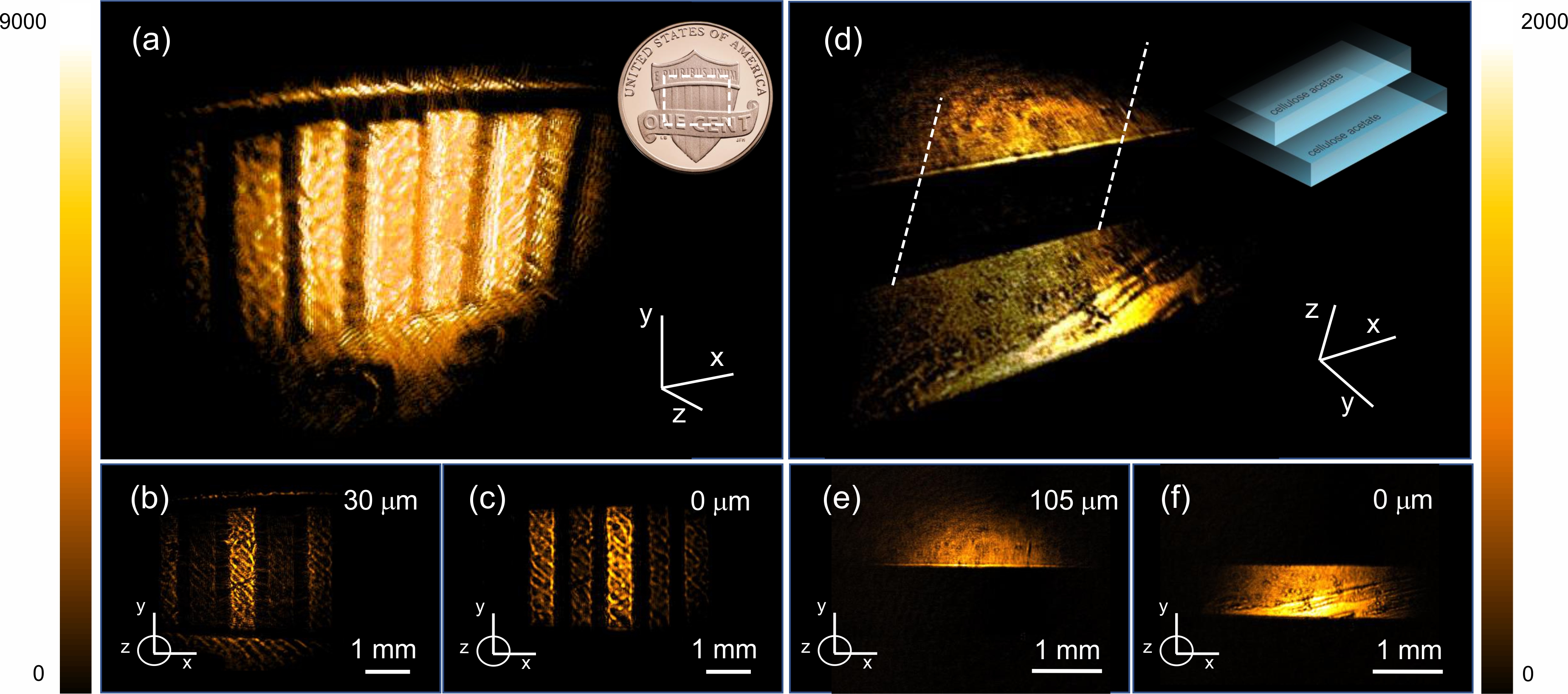}
    \caption{(a, b, c) Tomographic imaging of the structured metal surface of a one cent US coin (Union Shield). (a) 3D reconstruction, (b) and (c) are frames measured at height $h=30~\mu$m and $h=0~\mu$m, respectively. (d, e, f) Tomographic imaging of stacked cellulose acetate sheets, a weakly reflecting polymer structure. (d) 3D reconstruction, (e) and (f) are 2D frames taken at top of each sheet ($\Delta h=105~\mu$m). Total 3D scan time is 1 second.}
    \label{fig:fig2}
\end{figure}
We next study the 3D imaging capabilities of the wide-field NTA method by controlling the time delay between the MIR and NIR pulses. Figure \ref{fig:fig2}a depicts a 3D image of a one cent US coin. This reconstruction is comprised of individual 2D wide-field images acquired at 100~ms/frame using 10 steps along the axial dimension, corresponding to a $\sim75~\mu$m sample height and an effective total acquisition time of 1s. Two optical slices of the data stack are shown in Figure \ref{fig:fig2}b and \ref{fig:fig2}c, depicting the rectangular pillars of the Union Shield at different heights. The axial difference between these 2D layers is 30~$\mu$m, which corroborates the height of the structures as revealed with confocal microscopy (\href{https://www.chem.uci.edu/~dmitryf/images/Supplementary Information.pdf}{{\color{blue}Supplementary}} Figure SF3). 

Whereas reflection/scattering off metal/air interfaces allows high contrast imaging, in Figure \ref{fig:fig2}c we show that NTA-enabled detection also permits tomographic imaging of materials with a refractive index much closer to that of air. For this purpose, we used a polymer structure, consisting of a two-step ladder comprised of two 100~$\mu$m thick cellulose acetate sheets. The image reveals a $\sim105~\mu$m step size of the ladder, which closely matches the actual sheet thickness. The small difference between these values is attributed to flatness variations of the sheets, causing slight variations in the step size.

\section{\label{sec:third}3D imaging through transparent and highly absorbing media}
\begin{figure}
    \centering
    \includegraphics[width=14cm]{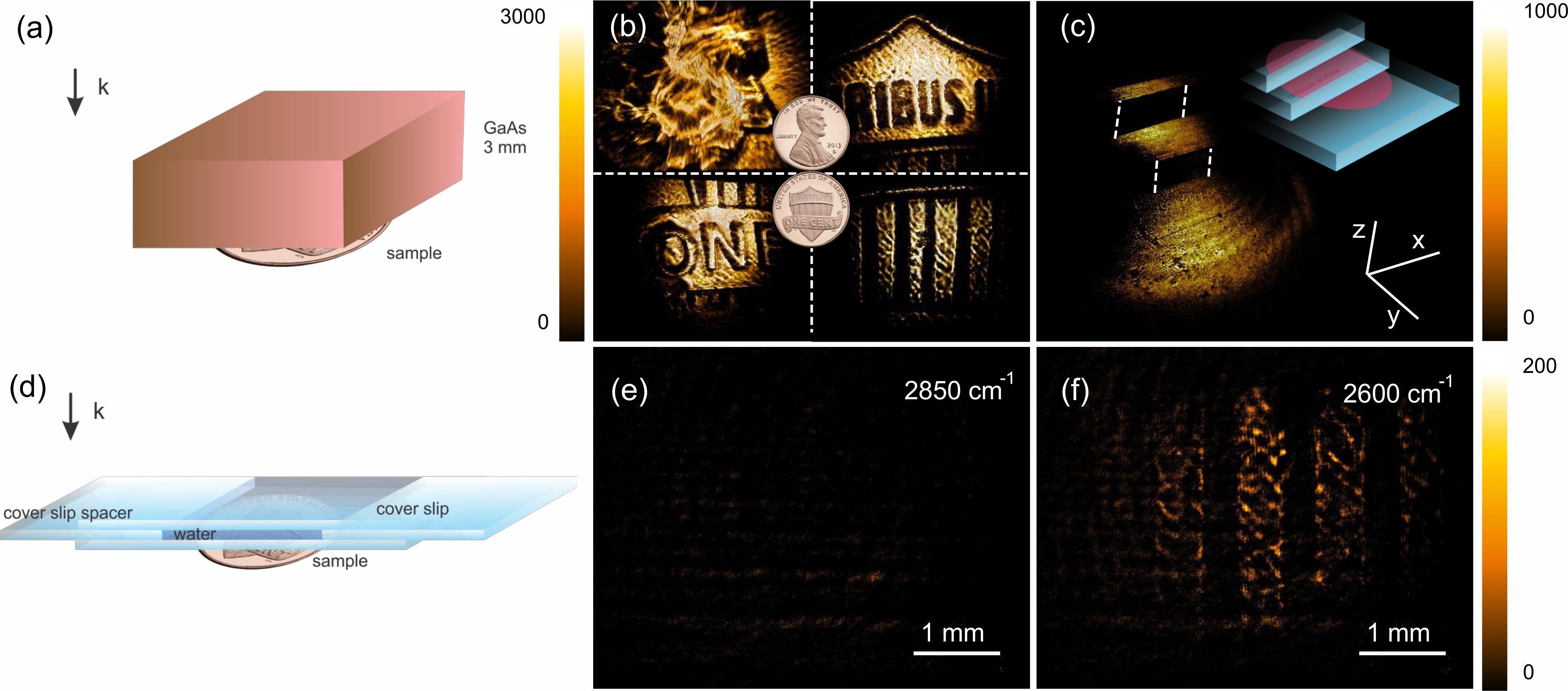}
    \caption{Sketch of penetration experiment arrangement through (a) 3~mm-thick GaAs wafer and (d) $190~\mu$m water layer. (b) 3D reconstruction of one cent US coin (Union Shield) through 3~mm GaAs wafer. (c) Tomographic imaging of stacked cellulose acetate sheets through 3~mm GaAs wafer. Imaging of one cent US coin (Union Shield) through $190~\mu$m water layer ($380~\mu$m in double pass) at (e) 2850~cm$^{-1}$ and (f) 2600~cm$^{-1}$.}
    \label{fig:fig3}
\end{figure}

Compared to tomographic imaging in the visible range, MIR-based tomography benefits from reduced light scattering in various solids, permitting imaging through thicker materials. To illustrate this point, we perform tomographic imaging of a coin hidden behind a 3~mm GaAs wafer, shown schematically in Figure \ref{fig:fig3}a. Although GaAs shows minimal absorption at the 2850~cm$^{-1}$ energy of the MIR pulse, reflection at the wafer’s top and bottom surfaces reduces the overall transmission of MIR light by 75\% in the double pass configuration (\href{https://www.chem.uci.edu/~dmitryf/images/Supplementary Information.pdf}{{\color{blue}Supplementary}} Figure SF4). Figure \ref{fig:fig3}b shows that, despite these losses, tomographic images of the coin’s features can still be clearly distinguished. Moreover, the weak reflection off a three-step cellulose acetate ladder provides sufficient back-scattered light for collecting a 3D image, depicted in Figure \ref{fig:fig3}c, even though it is covered by a visibly opaque material of high refractive index.
The detection sensitivity afforded by fs-NTA on the CCD camera also enables detection of objects placed under strongly absorbing materials. In Figure \ref{fig:fig3}d, we covered the coin with a 190~$\mu$m layer of deionized water. Water displays a strong MIR absorption due to the OH-stretching modes, which peaks in the 3000-3500 cm$^{-1}$ range, with broad wings that extend to lower energies beyond 2600 cm$^{-1}$. In Figure 3e, the coin is visualized when the MIR energy is set to 2850 cm$^{-1}$, showing low signal due to the strong water absorption. At this setting, a double pass through the water layer amounts to a MIR intensity loss of $\rm{OD}>4$ (\href{https://www.chem.uci.edu/~dmitryf/images/Supplementary Information.pdf}{{\color{blue}Supplementary}} Figure SF5). When tuned to 2650 cm$^{-1}$, however, a two-dimensional image of the coin surface can be observed under the water at a 100~ms exposure time, despite a MIR transmission loss of $\rm{OD}>2$. Here, the much lower signal levels necessitate subtraction of the DTA background ($\sim200$~counts), yet the results underline that fs-NTA detection is sensitive enough to retrieve MIR images even in the presence of thick water layers. 
\begin{figure}
    \centering
    \includegraphics[width=14cm]{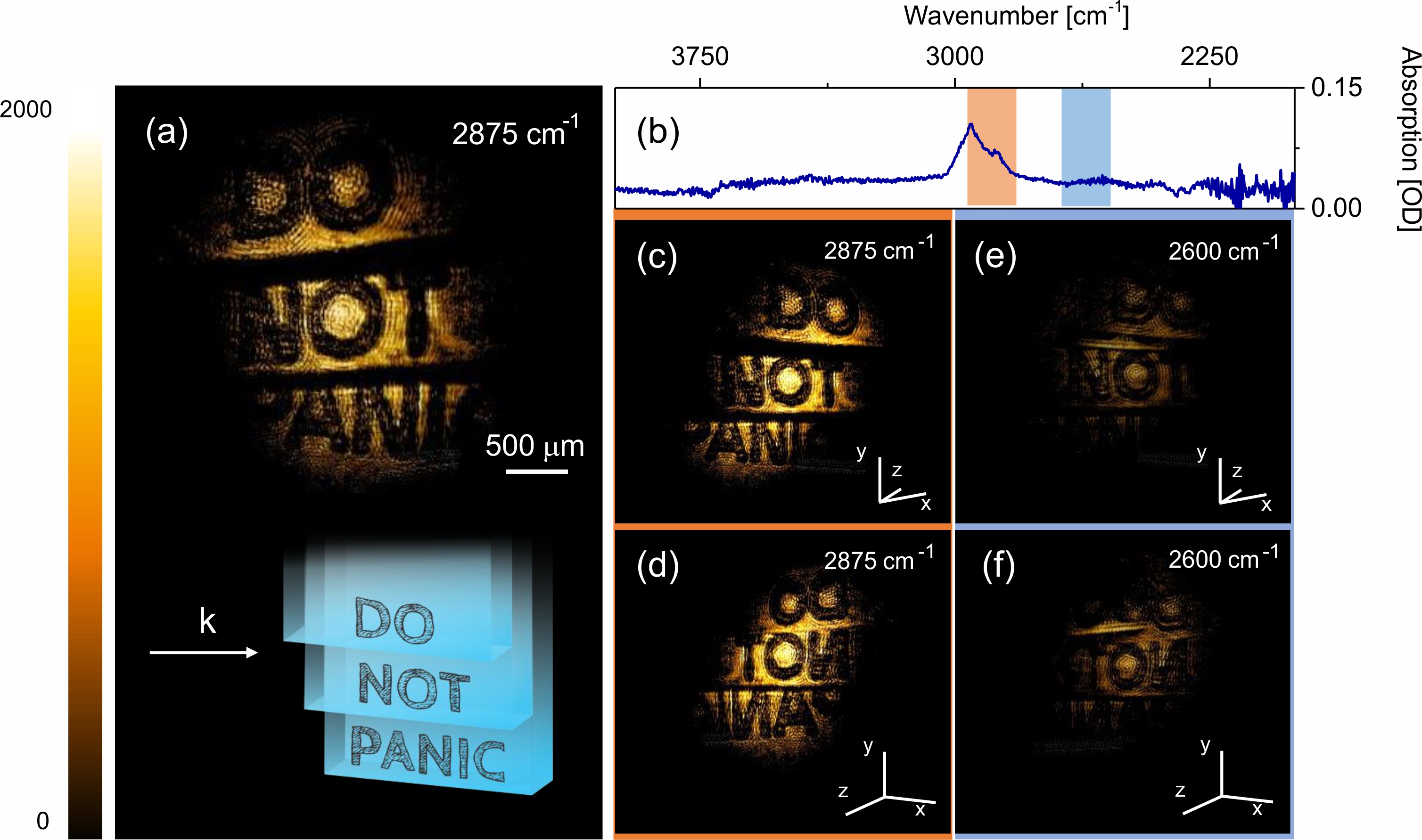}
    \caption{3D imaging of stacked cellulose acetate sheets with printed letters. (a) 3D reconstruction of the structure. (b) FTIR transmission spectrum of cellulose acetate. Rectangles represent Gaussian pulse width of $\sim150~\rm{cm}^{-1}$. (c) and (d) 3D imaging at 2875~cm$^{-1}$, (e) and (f) 3D imaging at 2600~cm$^{-1}$. Total image acquisition time is 1~s.}
    \label{fig:fig4}
\end{figure}

\section{\label{sec:fourth}Chemical-sensitive tomographic imaging}
One of the most promising aspects of MIR tomographic imaging is the possibility to generate chemically selective images of 3D objects. Using fs-NTA detection, we illustrate this principle by producing 3D images of the cellulose acetate ladder at different vibrational energies, shown in Figure \ref{fig:fig4}a. To enhance contrast, letters have been printed on each layer using black ink. The FTIR spectrum of cellulose acetate is plotted in Figure 4b. We may expect strong reflection off cellulose acetate when tuning to the red side of the CH-stretching vibration, where the real part of the complex refractive index displays a local maximum. When tuning to 2875 cm$^{-1}$, near the peak of the CH-stretching band, a bright tomographic image is obtained, as illustrated by the two projections in Figures \ref{fig:fig4}c and \ref{fig:fig4}d. At lower energies away from the resonance, the refractive index is expected to decrease, resulting in reduced reflection off the interface of the cellulose acetate material. This is indeed observed in Figures \ref{fig:fig4}e and \ref{fig:fig4}f, where the 3D image acquired at 2600~cm$^{-1}$ is now significantly less bright compared to the near-resonance condition at 2875~cm$^{-1}$. 
\begin{figure}
    \centering
    \includegraphics[width=14cm]{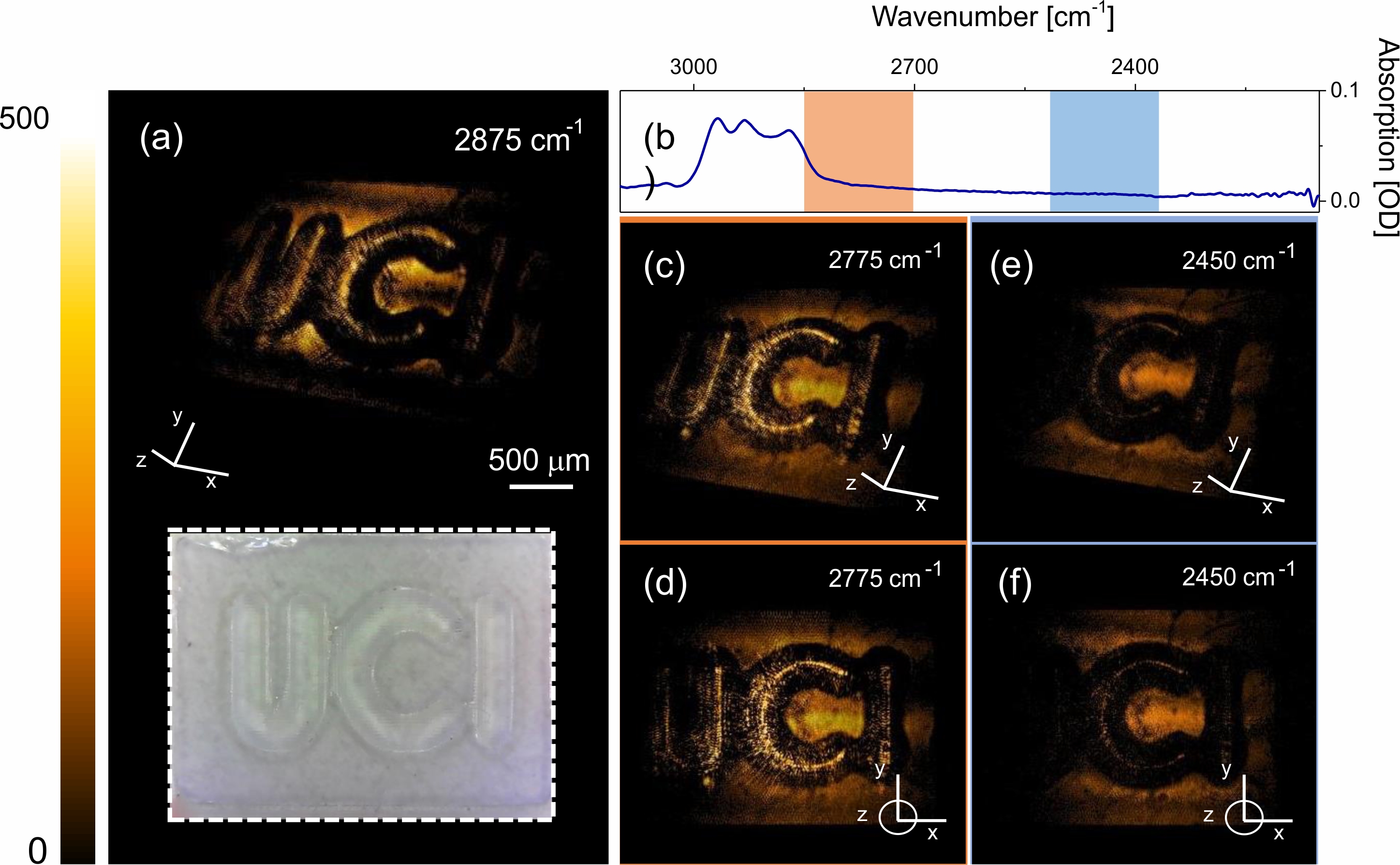}
    \caption{3D imaging of a resin structure manufactured through projection-based photolithography technique. (a) 3D reconstruction of resin structure. (b) FTIR absorption spectrum of the resin. Rectangles represent Gaussian pulse width of $\sim150~\rm{cm}^{-1}$. (c) and (d) 3D imaging at 2775~cm$^{-1}$, (e) and (f) 3D imaging at 2450~cm$^{-1}$. Structure height is $\sim50~\mu\rm{m}$. Total image acquisition time is 1~s.}
    \label{fig:fig5}
\end{figure}
The reduced reflection at 2600 cm$^{-1}$ enhances light penetration in the sample, which allows collection of signal contributions from lower lying interfaces. As shown in \href{https://www.chem.uci.edu/~dmitryf/images/Supplementary Information.pdf}{{\color{blue}Supplementary}} Figure SF6, signals from the buried interfaces (surface 2’ and 3’ in Figure SF6) are clearly observed in the off-resonance condition. Since the optical path length is determined by the refractive index of the material, the apparent depth of the buried interfaces differs from that of the corresponding air-exposed interfaces (surface 2 and 3 in Figure SF6).  From this time/optical path difference, we estimate the refractive index of cellulose acetate to be ~1.5 at 2600 cm$^{-1}$, resulting in a 4\% Fresnel reflection at the sheet/air interface. 

We next show 3D images of polymer structures fabricated with a projection-based photolithography technique (see Methods). A visible image of the structure is shown in the inset of Figure \ref{fig:fig5}a, and the relevant part of the FTIR spectrum of the polymer is given in Figure 5b. We observe increased signals from the structure’s top surface when the MIR energy is tuned into near-resonance with the material’s CH-stretching vibrational mode (Figures \ref{fig:fig5}c and \ref{fig:fig5}d), and lower signals when the MIR energy is tuned of off-resonance (Figures 5e and 5f). Due to the shape of the structure, significant light scattering occurs at angles beyond the collection NA of the imaging system, producing darker regions at curved surfaces. 

\begin{figure}
    \centering
    \includegraphics[width=14cm]{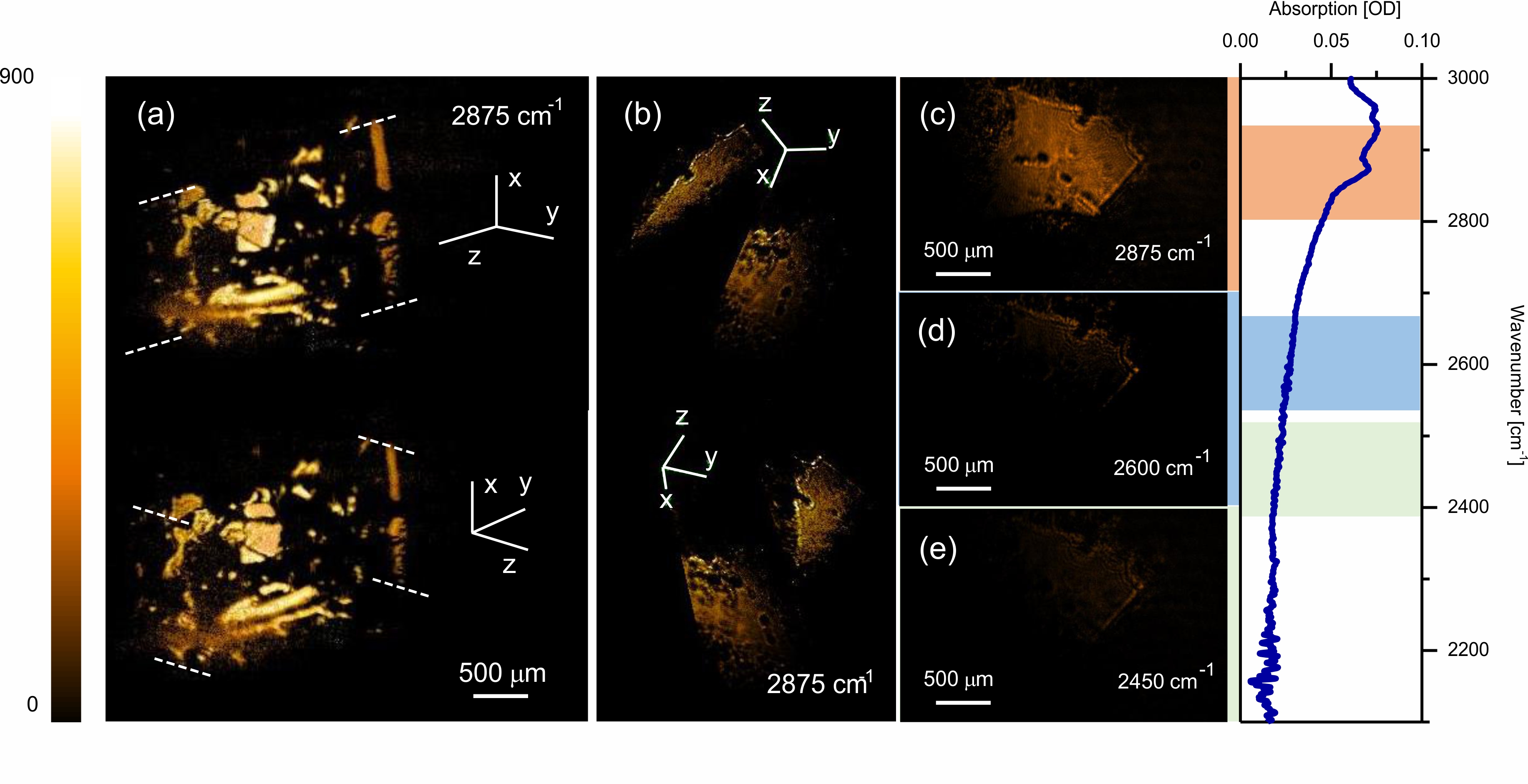}
    \caption{Imaging of different lysozyme crystals on mica glass. (a) 3D reconstruction of lysozyme crystal cluster at 2875 ~cm$^{-1}$. (b) 3D reconstruction of single crystal. 2D image of the crystal’s top face taken at (c) 2875~cm$^{-1}$, (d) 2600~cm$^{-1}$ and (f) 2450~cm$^{-1}$. FTIR absorption spectrum of lysozyme is shown on the far right.}
    \label{fig:fig6}
\end{figure}

Last, we perform MIR tomography of a hydrated protein crystal. In Figure \ref{fig:fig6}, we show a 3D reconstruction of lysozyme crystals. The lysozyme enzyme forms stable tetragonal crystals that can grow to millimeter scales. The structure visualized in Figure \ref{fig:fig6}a is composed of an aggregate of smaller crystals, while Figure \ref{fig:fig6}b shows a 3D image of a single crystal. The spectral dependence of the signal from a single surface is presented in Figure \ref{fig:fig6}c, \ref{fig:fig6}d and \ref{fig:fig6}e, confirming the chemical contrast encoded in the MIR light scattered onto the detector.

\section{\label{sec:fifth}Discussion}
In this work, we have discussed wide-field fs-NTA detection in Si cameras for enabling high-definition MIR tomographic imaging at high acquisition rates. Our technology is fundamentally different from existing OCT techniques that operate in the MIR. The most promising MIR-OCT approaches are currently based on Fourier-domain OCT (FD-OCT)\cite{Klein2017,Choma2003}, which accomplishes depth scans (A-scan) or cross-sectional scans (B-scans) on a time scale set by the spectral acquisition time of the detector. This detection strategy has been successfully translated to MIR-OCT using both classical\cite{Zorin2018} or quantum\cite{Vanselow2020} light, with A-scans acquired in just under 10~ms. However, in combination with lateral raster scanning, such imaging conditions translate into rather long total volume acquisition times of minutes/volume, extended further in duration by necessary referencing and post-processing  

In contrast, volumetric imaging through fs-NTA uses different mechanisms for capturing information along the lateral and axial dimensions. First, fs-NTA uses a wide-field detection approach, which allows detection of both lateral dimensions in a single shot. The massively parallel detection enabled by Mpx camera chips provides a dramatic gain of the effective acquisition rate. We have used an off-the-shelf CCD camera, with an effective readout time of 10~Hz. However, by using modern sCMOS cameras, which feature higher quantum yields and faster readout times, the image acquisition time can easily be improved with another two orders of magnitude. Second, the axial information in fs-NTA tomography is retrieved through a scan of the time delay between the MIR and the gate pulses. Scanning along this dimension can be accomplished with automated translation stages over 1~mm distances, with 1.5~$\mu$m (6 fs) repeatability and 2 ms response times. Hence, the total 3D acquisition time is limited primarily by the 2D image acquisition time of the camera. While in this work we report a total effective acquisition time of 1 s (10~frames/s, 10~axial steps) for the 3D data stack, it is evident that selection of a better camera would allow for volumetric imaging at much higher rates.

Compared to spectral interferometry approaches, volumetric imaging with fs-NTA is significantly more robust. First, it is not vulnerable to spectral shifts associated with small temperature changes of external nonlinear media, which may require recurring acquisition of multiple reference spectra.\cite{Vanselow2020,Wojtkowski2004} Second, in contrast to up-conversion techniques, fs-NTA mapping avoids the practical complication of phase-matching between the MIR and up-conversion pulses within the nonlinear medium.\cite{Junaid2018,Junaid2019,Israelsen2019}Wide-field fs-NTA requires no DTA-background subtraction nor multiple processing steps to retrieve the MIR signal. Instead, 2D images are acquired in single shot mode, allowing rapid and unimpeded axial scans for collecting 3D data stacks. For our current imaging conditions, the signal-to-noise ratio (SNR) of each frame is 72~dB for a 100~ms integration time, determined by using a gold mirror as the sample. The NTA to DTA background ratio is 34~dB, which is sufficiently high for enabling imaging under virtually background-free conditions. 

We have shown that MIR tomographic imaging enabled by fs-NTA exhibits chemical contrast based on vibrational resonances of the sample. Although the spectral resolution is limited by the bandwidth of the MIR pulse ($<150~\rm{cm}^{-1}$), the spectroscopic contrast imparted by the sample’s vibrational modes is clearly observed. It is non-trivial to achieve similar contrast with conventional FD-OCT methods, where spectroscopic imprints are difficult to retrieve in the presence of the strong interferometric modulations of the detected spectrum. Therefore, although MIR-OCT is recognized for its greater penetration depth, chemically selective imaging has remained a challenge. As our experiments show, chemical contrast appears rather naturally when the fs-NTA detection approach is used.

Finally, we note that our current work utilizes a light source based on a low repetition rate amplified laser system. Such amplified pulses, however, are not a prerequisite for the demonstrated imaging capabilities. We have previously shown that a detection sensitivity of only few fJ/pulse per pixel can be achieved with ps pulses derived from a high repetition rate picosecond light source.\cite{Knez2020} Similar experiments with shorter, femtosecond pulses will require only few aJ of MIR radiation on single camera pixel. Emerging developments in ultrafast fiber-based lasers promise to provide such high repetition rate fs pulses, underlining the potential of rapid fs-NTA tomography with affordable and compact light sources.

\section{\label{sec:sixth}Methods}

\subsection{Fourier Transform Infrared Spectroscopy}
MIR absorption of materials is measured with a commercial FTIR spectrometer (Jasco 4700) either in transmission mode or by using an ATR accessory equipped with a diamond crystal.

\subsection{MIR fs-NTA imaging system}
The imaging system is schematically depicted in Figure 1a and described in the manuscript. A 1~kHz amplified femtosecond laser system (Spitfire Ace, Spectra Physics) is used to seed two optical parametric amplifiers (Topas Prime, Light Conversion). One OPA is used as a source of NIR gate radiation at 1200~nm. The signal and idler pulses from the second OPA system are used in to generate MIR pulses through the process of difference frequency generation in a nonlinear medium. Both MIR and NIR pulses are re-combined on a 1~mm ZnSe window, which serves as dichroic mirror, after which the pulses are overlapped on a CCD camera (DR-328G-CO2-SIL Clara, Andor). Temporal overlap is controlled through a mechanical delay stage (GTS150, Newport). The MIR imaging system consist of two 100~mm CaF$_2$ lenses, resulting in a 1:1 imaging system with NA=0.015. The incident angle of the MIR beam on the sample is less than 5 degrees, resulting in an error for z-height determination of less than a percent.

\subsection{Polymer structure fabrication}
The computer-aided design models of the structures (SolidWorks) are virtually sliced into 2D layers with a slice thickness of $20~\mu$m. Mask projection images are generated for each layer.\cite{Zhou2009} The exposure time of each layer is adjusted based on the light intensity and the photosensitivity of the printing material (ranging from 5~s to 8~s) to improve the fabrication accuracy. The UV-curing photopolymer resin from Elegoo Inc. is used for structure fabrication due to its desired IR property. The resin is used directly without modification. 

In the projection-based stereolithography process, the photocurable resin is deposited on the surface of a transparent resin tank. To generate the 2D patterned light beam, 405~nm wavelength light is reflected by a digital micromirror device (DMD) comprised of a $1920\times1080$ array of micromirrors, and the brightness of each pixel in the projected light beam is controlled by adjusting the angle of the corresponding micromirror in the DMD. 

\subsection{Confocal imaging of coin structure}
The 3D-images are acquired with a Leica SP8 Dive microscope operated in the reflectance confocal imaging mode using a 532 nm light source and a $10\times$, 0.3~NA objective. The images are acquired as z-stacks of mosaics (adjacent field-of-views stitched together). The area of each en-face mosaic frame is $13.5\times9.3$~mm, the distance between the frames in the z-stack is $5~\mu$m.

\subsection{Lysozyme crystals growing and handling}
Hen egg white lysozyme is purchased from Fisher Scientific (ICN19530325). The lyophilized powder is dissolved to 20~mg/mL in 100~mM sodium acetate pH 4.5. Batch crystallization is performed with the lysozyme solution in a 1:1 ratio with 1 M NaCl in 100~mM sodium acetate pH 4.5.

\section{Acknowledgements}
R.W.M. and B.N.-B. would like to acknowledge NSF BMAT DMR-2002837 and the NSF GRFP. Y.C. would like to acknowledge the support of NSF grant CMMI 1151191. A.D., A.F. and M.B. would like to acknowledge NIH award 1S10OD028698. E.O.P. acknowledges NIH grant R01GM132506 and NSF grant CMMI 1905582. DAF thanks Yulia Davydova for help with nonlinear experiments and
support. We thank the Laser Spectroscopy Labs.

\bibliography{bibliography}

\end{document}